\documentclass[prl,aps,twocolumn]{revtex4}

\usepackage{psfrag}
\usepackage{graphicx}
\usepackage{dcolumn}
\usepackage{latexsym,amsfonts}
\usepackage{bm}
\usepackage{amssymb}
\begin{document}

\title{On Continuum-State and Bound-State $\beta^-$--Decay Rates of the
Neutron}

\author{M. Faber$^{a}$\thanks{E-mail: faber@kph.tuwien.ac.at},
A. N. Ivanov${^a}$, V. A. Ivanova$^{b}$, J. Marton$^{c}$,\\
M. Pitschmann${^a}$, A. P. Serebrov${^d}$, N. I. Troitskaya$^{b}$,
M. Wellenzohn$^{a}$} \affiliation{${^a}$Atominstitut der
\"Osterreichischen Universit\"aten, Technische Universit\"at Wien,
Wiedner Hauptstrasse 8-10, A-1040 Wien, Austria}\affiliation{ $^b$
State Polytechnic University of St. Petersburg, Polytechnicheskaya 29,
195251, Russian Federation} \affiliation{${^c}$Stefan Meyer Institut
f\"ur subatomare Physik \"Osterreichische Akademie der Wissenschaften,
Boltzmanngasse 3, A-1090, Wien, Austria}\affiliation{ $^d$Petersburg
Nuclear Physics Institute, 188300 Gatchina, Orlova roscha 1, Russian
Federation}

 \email{ivanov@kph.tuwien.ac.at}

\date{\today}

\begin{abstract}
We analyse the continuum-state and bound-state $\beta^-$--decay rates
of the neutron. For the calculation of theoretical values of the decay
rates we use the new value for the axial coupling constant $g_A =
1.2750(9)$, obtained recently by H. Abele (Progr. Part. Nucl. Phys.,
{\bf 60}, 1 (2008)) from the fit of the experimental data on the
neutron spin-electron correlation coefficient of the electron energy
spectrum of the continuum-state $\beta^-$--decay of the neutron. We
take into account the contribution of radiative corrections and the
scalar and tensor weak couplings. We define correlation coefficients
of the electron energy spectrum in terms of axial, scalar and tensor
coupling constants.  Using recent precise experimental data for the
lifetime of the neutron and correlation coefficients we estimate the
scalar and tensor weak coupling constants.  The bound-state
$\beta^-$--decay rates of the neutron we calculate as functions of
axial, scalar and tensor weak coupling constants. We show that
dominantly the neutron decays into hydrogen in the hyperfine states
with total angular momentum $F = 0$.  The calculated angular
distributions of the probabilities of the bound-state
$\beta^-$--decays of the polarised neutron can be used for the
experimental measurements of the bound-state $\beta^-$--decays into
the hyperfine states with a total angular momentum $F = 1$. 
\end{abstract}

\maketitle

\section{Introduction}

The continuum-state $\beta^-$--decay of the neutron $n \to p + e^- +
\tilde{\nu}_e$ is experimentally well measured \cite{NDR1,NDR2} (see
also \cite{NDR3} and \cite{PDG08}) and investigated theoretically
\cite{JJ57,NS06}. A theoretical analysis of the bound-state
$\beta^-$--decay rate has been carried out in \cite{BS1,BS2}. Recently
\cite{WS1,WS2} Schott {\it et al.} have reported new experimental data
on the bound-state $\beta^-$--decay of the neutron $n \to {\rm H} +
\tilde{\nu}_e$.

In this paper we recalculate the continuum-state $\beta^-$--decay rate
of the neutron, the electron energy spectrum and angular distribution
taking into account the contributions of $V - A$, scalar $S$ and
tensor $T$ weak interactions and radiative corrections
\cite{RC1,RC2}. Such a recalculation is required by the new precise
experimental data on the lifetime of the neutron $\tau_{\beta^-_c} =
878.5(8)\,{\rm s}$ \cite{NDR1} and the value of the axial coupling
constant $g_A = 1.2750(9)$ \cite{NDR3}. Using recent experimental data
on the lifetime of the neutron \cite{NDR1} and correlation
coefficients \cite{NDR3} we estimate the scalar $g_S$ and tensor $g_T$
coupling constants.  For the experimental analysis of the
contributions of the scalar and tensor weak interactions we give
angular distributions of the probabilities of the bound-state
$\beta^-$--decay rates of the polarised neutron. The calculation of
the bound-state $\beta^-$--decay rates we use the technique applied to
the analysis of the weak decays of the H--like, He--like and bare
heavy ions and mesic hydrogen in \cite{Ivanov1}--\cite{Faber2}.  In
the Conclusion we discuss the obtained results. In the Appendix we
discuss the radiative corrections.

\section{$V-A$ weak hadronic interactions}

The Hamiltonian of weak interaction we take in the form
\cite{Ivanov1}--\cite{Faber2}
\begin{eqnarray}\label{label1}
\hspace{-0.3in}{\cal H}_W(x) &=&
\frac{G_F}{\sqrt{2}}\,V_{ud}\,[\bar{\psi}_p(x)\gamma_{ \mu}(1 -
g_A\gamma^5)\psi_n(x)]\nonumber\\
\hspace{-0.3in}&&\times\,[\bar{\psi}_e(x)\gamma^{\mu}(1 -
\gamma^5)\psi_{\nu_e}(x)],
\end{eqnarray}
where $G_F = 1.1664\times 10^{-11}\,{\rm MeV}^{-2}$ is the Fermi weak
constant, $V_{ud}$ and $g_A$ are the CKM matrix element and the axial
coupling constant \cite{PDG08}, $\psi_p(x)$, $\psi_n(x)$, $\psi_e(x)$
and $\psi_{\nu_e}(x)$ are operators of interacting proton, neutron,
electron and anti-neutrino, respectively.

For numerical calculations we will use the most precise values
$|V_{ud}| = 0.97419(22)$ \cite{PDG08} and $g_A = 1.2750(9)$
\cite{NDR3}, where $g_A = 1.2750(9)$ has been obtained from the fit of
the neutron spin--electron correlation coefficient $A^{\exp} = -
0.11933(34)$, defined in terms of the axial coupling $g_A$ in
Eq.(\ref{label32}), of the electron energy spectrum for the
continuum-state $\beta^-$--decay of the neutron \cite{NDR3}.

The value of the CKM matrix element $|V_{ud}| = 0.97419(22)$
\cite{PDG08} agrees well with $|V_{ud}| = 0.9738(4)$ \cite{NDR3,NDR5},
measured from the superallowed $0^+ \to 0^+$ nuclear
$\beta^-$--decays, which are pure Fermi transitions \cite{NDR5}. It
satisfies also well the unitarity condition $|V_{ud}|^2 + |V_{us}|^2 +
|V_{ub}|^2 = 1.0000(6)$ for the CKM matrix elements \cite{PDG08}.

The $\mathbb{T}$--matrix of weak interactions is equal to
\begin{eqnarray}\label{label2}
\mathbb{T} = -\int d^4x\,{\cal H}_W(x).
\end{eqnarray}
The amplitudes of the continuum-state and bound-state
$\beta^-$--decays of the  neutron are defined by the matrix
elements of the $\mathbb{T}$--matrix
\begin{eqnarray}\label{label3}
\langle \tilde{\nu}_e e^- p|\mathbb{T}|n\rangle &=&
(2\pi)^4\delta^{(4)}(k_{\tilde{\nu}_e} + k_e + k_p - k_n)\nonumber\\
&&\times\,M(n \to p + e^- + \tilde{\nu}_e),\nonumber\\ \langle
\tilde{\nu}_e H|\mathbb{T}|n\rangle &=&
(2\pi)^4\delta^{(4)}(k_{\tilde{\nu}_e} + k_{\rm H} - k_n)\nonumber\\
&&\times\,M(n \to k_H + \tilde{\nu}_e).
\end{eqnarray}
The amplitudes $M(n \to p + e^- + \tilde{\nu}_e)$ and $M(n \to k_H +
\tilde{\nu}_e)$ of the decays are 
\begin{eqnarray}\label{label4}
M(n \to p + e^- + \tilde{\nu}_e) &=& - \langle \tilde{\nu}_e
e^-p|{\cal H}_W(0)|n\rangle ,\nonumber\\ M(n \to k_H + \tilde{\nu}_e)
&=& - \langle \tilde{\nu}_e {\rm H}|{\cal H}_W(0)|n\rangle ,
\end{eqnarray}
where $k_a$ with $a = \tilde{\nu}_e, e, p,{\rm H}$ and $n$ are the
4--momenta of interacting particles.

\section{Bound-state and continuum-state $\beta^-$--decay rates of  
neutron in $V - A$ theory of weak interactions}

In the final state of the bound-state $\beta^-$--decay of the neutron
hydrogen can be produced only in the $ns$--states, where $n$ is a {\it
principal} quantum number $n = 1,2,\ldots $ \cite{Faber1,Faber2}. The
contribution of the excited $n\ell$-state with $\ell > 0$ is
negligibly small. Due to hyperfine interactions \cite{BS57,HFS1}
hydrogen can be in two hyperfine states $(ns)_F$ with $F = 0$ and $F =
1$

The wave function of hydrogen ${\rm H}$ in the
$ns$--state we take in the form \cite{IV2}--\cite{IV5} 
\begin{eqnarray}\label{label5}
\hspace{-0.3in}&&|{\rm H}^{(ns)}(\vec{q}\,)\rangle =
 \frac{1}{(2\pi)^3}\sqrt{2 E_{\rm H}(\vec{q}\,)}\nonumber\\
\hspace{-0.3in}&&\times\int \frac{d^3k_e}{\sqrt{2E_e(\vec{k}_e)}}
 \frac{d^3k_p}{\sqrt{2E_p(\vec{k}_p)}}\, \delta^{(3)}(\vec{q} -
 \vec{k}_e - \vec{k}_p)\nonumber\\
\hspace{-0.3in}&&\times \phi_{ns}\Big(\frac{m_p \vec{k}_e - m_e
 \vec{k}_p}{m_p + m_e}\Big) a^{\dagger}_{ns}(\vec{k}_e,\sigma_e)
 a^{\dagger}_p(\vec{k}_p,\sigma_p)|0\rangle,
\end{eqnarray}
where $E_{\rm H}(\vec{q}\,) = \sqrt{M^{\;2}_{\rm H} + \vec{q}^{\;2}}$
and $\vec{q}$ are the total energy and the momentum of hydrogen,
$M_{\rm H} = m_p + m_e + \epsilon_{ns}$ and $\epsilon_{ns}$ are the
mass and the binding energy of hydrogen ${\rm H}$ in the $(ns)_F$
hyperfine state; $\phi_{ns}(\vec{k}\,)$ is the wave function of the
$ns$--state in the momentum representation \cite{BS57} (see also
\cite{IV2}--\cite{IV5}). For the calculation of the bound state
$\beta^-$--decay rate we can neglect the hyperfine splitting of the
energy levels of the $ns$--states \cite{BS57,HFS1}.

For the amplitude of the bound-state $\beta^-$--decay we obtain the
following expression
\begin{eqnarray}\label{label6}
\hspace{-0.3in}&&M(n\to {\rm H}^{(ns)} + \tilde{\nu}_e) = G_F
V_{ud}\sqrt{2 m_n 2 E_{\rm H}2 E_{\tilde{\nu}_e}}\nonumber\\
\hspace{-0.3in}&&\times\int
\frac{d^3k}{(2\pi)^3}\,\phi^*_{ns}\Big(\vec{k} - \frac{m_e}{m_p +
m_e}\,\vec{q}\,\Big)\,\Big\{[\varphi^{\dagger}_e\chi_{_{\tilde{\nu}_e}}]\nonumber\\
\hspace{-0.3in}&&\times\,[\varphi^{\dagger}_p \varphi_n] - g_A\,
[\varphi^{\dagger}_e\vec{\sigma}\,\chi_{_{\tilde{\nu}_e}}]\cdot [
\varphi^{\dagger}_p\vec{\sigma}\,\varphi_n]\Big\},
\end{eqnarray}
where $\varphi_p$, $\varphi_n$, $\varphi_e$ and
$\chi_{_{\tilde{\nu}_e}}$ are spinorial wave functions of the proton,
neutron, electron and antineutrino.  The integral over $\vec{k}$ of
the wave function $\phi^*_{ns}(\vec{k}\,)$ defines the wave function
$\psi^*_{ns}(0)$ in the coordinate representation, equal to
$\psi^*_{ns}(0) = \sqrt{\alpha^3 m^3_e/n^3 \pi}$, where $m_e$ is the
electron mass and $\alpha = 1/137.036$ is the fine--structure
constant.  This gives
\begin{eqnarray}\label{label7}
\hspace{-0.3in}&&M(n\to {\rm H}^{(ns)} + \tilde{\nu}_e) = G_F
V_{ud}\sqrt{2 m_n 2 E_{\rm H}2 E_{\tilde{\nu}_e}}\nonumber\\
\hspace{-0.3in}&&\times\,\Big\{[\varphi^{\dagger}_e
\chi_{_{\tilde{\nu}_e}}][\varphi^{\dagger}_{p} \varphi_n] -
g_A[\varphi^{\dagger}_e\vec{\sigma}\,\chi_{_{\tilde{\nu}_e}}]\cdot [
\varphi^{\dagger}_p\vec{\sigma}\,\varphi_n] \Big\}\nonumber\\
\hspace{-0.3in}&& \times\,\psi^*_{(ns)_F}(0).
\end{eqnarray}
The bound-state $\beta^-$--decay rate of the  neutron is
\begin{eqnarray}\label{label8}
\hspace{-0.3in}&&\lambda_{\beta^-_b} = \frac{1}{2m_n}\int
\frac{1}{2}\sum^{\infty}_{n=1}\sum_{\sigma_n,\sigma_p,\sigma_e}\!\!\!\!\!|M(n\to
{\rm H}^{(ns)} + \tilde{\nu}_e)|^2\nonumber\\
\hspace{-0.3in}&&\times(2\pi)^4\delta^{(4)}(k_{\tilde{\nu}_e} + q -
p)\,\frac{d^3q}{(2\pi)^3 2 E_{\rm H}}\frac{d^3k_{\tilde{\nu}_e}}{(2\pi)^3 2
E_{\tilde{\nu}_e}}.
\end{eqnarray}
Summing over the {\it principal} quantum number and polarisations 
we get
\begin{eqnarray}\label{label9}
\lambda_{\beta^-_b} &=& (1 + 3 g^2_A)\,\zeta(3)\,
G^2_F|V_{ud}|^2\frac{\alpha^3 m^3_e}{\pi^2}\nonumber\\
&\times&\sqrt{(m_p + m_e)^2 + Q^2_{\beta^-_c}}\,
\frac{Q^2_{\beta^-_c}}{m_n},
\end{eqnarray}
where $\zeta(3) = 1.202$ is the Riemann function, coming from the
summation over the {\it principal} quantum number $n$, and
$Q_{\beta^-_c}$ is the $Q$--value of the continuum-state
$\beta^-$--decay of the  neutron equal to
\begin{eqnarray}\label{label10}
 Q_{\beta^-_c} = \frac{m^2_n - (m_p + m_e)^2}{2 m_n} = 0.782\,{\rm
   MeV}.
\end{eqnarray}
In the literature \cite{BS1,BS2} the bound-state $\beta^-$--decay rate
of the neutron is defined relative to the continuum-state
$\beta^-$--decay rate of the neutron. 

The theoretical value of the continuum-state $\beta^-$--decay rate of
the neutron is
\begin{eqnarray}\label{label11}
  \lambda_{\beta^-_c} &=& (1 + 3 g^2_A)\, \frac{G^2_F |V_{ud}|^2}{2
    \pi^3}\,f(Q_{\beta^-_c}, Z = 1) = \nonumber\\ &=&1.0931(14)\times
    10^{-3}\,{\rm s}^{-1},
\end{eqnarray}
where the error of the decay rate is fully defined by the experimental
error of the axial coupling constant $g_A = 1.2750(9)$ and the CKM
matrix element $|V_{ud}| = 0.97419(22)$. The numerical value of the
continuum-state $\beta^-$--decay rate of the  neutron is
calculated for the experimental masses of the interacting particles
\cite{PDG08} and the Fermi integral $f(Q_{\beta^-_c}, Z = 1)$ equal to
\begin{eqnarray}\label{label12}
\hspace{-0.3in}&&f(Q_{\beta^-_c}, Z = 1) = \int^{Q_{\beta^-_c} +
m_e}_{m_e} (Q_{\beta^-_c} + m_e - E_e )^2\nonumber\\
\hspace{-0.3in}&& \times\, E_e \sqrt{E^2_e - m^2_e}\,F(E_e, Z =
1)\,dE_e = \nonumber\\
\hspace{-0.3in}&& = \int^{Q_{\beta^-_c} + m_e}_{m_e} \frac{2\pi\alpha
E^2_e (Q_{\beta^-_c} + m_e - E_e )^2 }{\displaystyle 1 - e^{\textstyle
-\,2\pi \alpha E_e /\sqrt{E^2_e - m^2_e}}}\,dE_e = \nonumber\\
\hspace{-0.3in}&& = 0.0588\,{\rm MeV}^5
\end{eqnarray}
and the Fermi function \cite{HS66}
\begin{eqnarray}\label{label13}
 \hspace{-0.3in}&& F(E_e, Z = 1) = \nonumber\\
\hspace{-0.3in}&& =\frac{2\pi\alpha E_e}{\sqrt{E^2_e -
m^2_e}}\,\frac{1}{\displaystyle 1 - e^{\textstyle -\,2\pi \alpha
E_e/\sqrt{E^2_e - m^2_e}}}.
\end{eqnarray}
The theoretical value of the lifetime of the  neutron is
$\tau_{\beta^-_c} = 914.8(1.2)\,{\rm s}$, defined by $\tau_{\beta^-_c}
= 1/\lambda_{\beta^-_c}$.

Taking into account the radiative corrections \cite{RC1,RC2} (see the
Appendix), the theoretical value of the lifetime reduces to
$\tau^{(\gamma)}_{\beta^-_c} = 880.4.(1.1)\,{\rm s}$. It agrees well
with the experimental $\tau^{\exp}_{\beta^-_c} = 878.5(8)\,{\rm s}$
\cite{NDR1}. The theoretical value of the lifetime
$\tau^{(\gamma)}_{\beta^-_c} = 880.4(1.1)\,{\rm s}$ differs from the
world averaged experimental value $\tau^{\exp}_{\beta^-_c} =
885.7(8)\,{\rm s}$ \cite{PDG08} by a few seconds $(-\,5.3\pm
1.4)\,{\rm s}$.

For the ratio $R_{b/c} = \lambda_{\beta^-_b}/
\lambda^{(\gamma)}_{\beta^-_c}$ of the bound-state and continuum-state
$\beta^-$--decay rates of the  neutron we get the following
expression
\begin{eqnarray}\label{label14}
\hspace{-0.27in} &&R_{b/c}= \zeta(3) 2\pi\frac{\alpha^3m^3_e
 Q^2_{\beta^-_c}}{m_n}\frac{\sqrt{(m_p + m_e)^2 +
 Q^2_{\beta^-_c}}}{f^{(\gamma)}(Q_{\beta^-_c}, Z = 1)}=\nonumber\\ 
\hspace{-0.27in} &&=  3.92\times 10^{-6},
\end{eqnarray}
where the Fermi integral $f^{(\gamma)}(Q_{\beta^-_c}, Z = 1) =
0.0611\,{\rm MeV^5}$ is calculated in the Appendix
Eq.(\ref{labelA.1}).  Our value for the ratio of the decay rates
agrees with the results obtained in \cite{BS1} (see also
\cite{WS1,WS2}): $R_{b/c} = 4.20\times 10^{-6}$. 

In spite of such a success of the $V - A$ theory of weak interactions
for the description of the continuum-state $\beta^-$--decay rate of
the neutron, in the next section we take into account the
contributions of scalar $S$ and tensor $T$ weak interactions of
baryons and leptons and estimate the values of the scalar and tensor
coupling constants \cite{JJ57,NS06}.

\section{Continuum-state and bound-state $\beta^-$--decay rates of  
neutron in $V - A$, scalar and tensor theory of weak interactions}
 
In this section we consider the continuum-state and bound-state
$\beta^-$--decays of the neutron by taking into account scalar and
tensor weak interactions \cite{JJ57,NS06}. The effective
low--energy Hamiltonian of these interactions can be taken in the
following form
\begin{eqnarray}\label{label15}
\hspace{-0.3in}&&\tilde{{\cal H}}_W(x) = 
\frac{G_F}{\sqrt{2}}\,V_{ud}\,\Big\{g_S\,[\bar{\psi}_p(x)
\psi_n(x)]\nonumber\\
\hspace{-0.3in}&&\times\,[\bar{\psi}_e(x)(1 -
\gamma^5)\psi_{\nu_e}(x)] +
\frac{1}{2}\,g_T\,[\bar{\psi}_p(x)\sigma_{\mu\nu}\gamma^5\psi_n(x)]\nonumber\\
\hspace{-0.3in}&&\,[\bar{\psi}_e(x) \sigma^{\mu\nu}(1 -
\gamma^5)\psi_{\nu_e}(x)]\Big\},
\end{eqnarray}
where $g_S$ and $g_T$ are constants of scalar and tensor weak
interactions and $\sigma_{\mu\nu} =
\frac{1}{2}(\gamma_{\mu}\gamma_{\nu} - \gamma_{\nu}\gamma_{\mu})$ is
the Dirac matrix.

In the non--relativistic approximation for the neutron and the proton
 the contribution of the scalar and tensor weak interactions to the
 amplitude of the continuum-state $\beta^-$--decay is
\begin{eqnarray}\label{label16}
\hspace{-0.3in}&&\tilde{M}(n \to p + e^- + \tilde{\nu}_e) = -
\frac{G_F}{\sqrt{2}}\,V_{ud}\,\sqrt{ 4m_p m_n}\nonumber\\
\hspace{-0.3in}&&\times \Big\{g_S[\bar{u}_e(\vec{k}_e,\sigma_e)\,(1 -
\gamma^5)v_{\tilde{\nu}_e}(\vec{k}_{\tilde{k}_e},+
\frac{1}{2})][\varphi^{\dagger}_p
\varphi_n]\nonumber\\
\hspace{-0.3in}&& + g_T [\bar{u}_e(\vec{k}_e,\sigma_e)\vec{\alpha}\,(1
- \gamma^5)v_{\tilde{\nu}_e}(\vec{k}_{\tilde{\nu}_e},+
\frac{1}{2})][\varphi^{\dagger}_p
\vec{\sigma}\,\varphi_n]\Big\},\nonumber\\
\hspace{-0.3in}&&
\end{eqnarray}
where $\vec{\alpha} = \gamma^0\vec{\gamma}$ is the Dirac matrix.

The total amplitude of the continuum-state $\beta^-$--decay of the
neutron, containing the contributions of $V - A$, $S$ and $T$
interactions, is
\begin{eqnarray}\label{label17}
\hspace{-0.3in}&&M(n \to p + e^- + \tilde{\nu}_e) = -
\frac{G_F}{\sqrt{2}}\,V_{ud}\,\sqrt{ 4 m_p m_n}\nonumber\\
\hspace{-0.3in}&&\times
\Big\{[\bar{u}_e(\vec{k}_e,\sigma_e)\,(\gamma^0 + g_S)\,(1 -
\gamma^5)v_{\tilde{\nu}_e}(\vec{k}_{\tilde{k}_e},+
\frac{1}{2})]\nonumber\\ \hspace{-0.3in}&&\times\,[\varphi^{\dagger}_p
\varphi_n]\nonumber\\
\hspace{-0.3in}&& + [\bar{u}_e(\vec{k}_e,\sigma_e)(g_A \gamma^0 +
g_T)\,\vec{\alpha}\,(1 -
\gamma^5)v_{\tilde{\nu}_e}(\vec{k}_{\tilde{\nu}_e},+
\frac{1}{2})]\nonumber\\
\hspace{-0.3in}&&\cdot [\varphi^{\dagger}_p \vec{\sigma}\,\varphi_n]
\Big\}.
\end{eqnarray}
The theoretical value of the continuum-state $\beta^-$--decay rate of
the  neutron, accounting for the contributions of scalar and
tensor weak interactions, is
\begin{eqnarray}\label{label18}
\hspace{-0.3in}&&\tilde{\lambda}_{\beta^-_c} = \frac{G^2_F
    |V_{ud}|^2}{2 \pi^3}\nonumber\\
\hspace{-0.3in}&&\times\,\Big\{((1 + 3 g^2_A) + (g^2_S + 3
g^2_T))\, f^{(\gamma)}(Q_{\beta^-_c}, Z = 1)\nonumber\\
\hspace{-0.3in}&& + 2 (g_S + 3 g_A g_T)\,\tilde{f}^{(\gamma)}(Q_{\beta^-_c}, Z =
1)\Big\},
\end{eqnarray}
where $\tilde{f}(Q_{\beta^-_c}, Z = 1)$ is the Fermi integral equal to 
\begin{eqnarray}\label{label19}
\hspace{-0.3in}&&\tilde{f}(Q_{\beta^-_c}, Z = 1) = \nonumber\\
\hspace{-0.3in}&& = \int^{Q_{\beta^-_c} + m_e}_{m_e} \frac{2\pi\alpha
m_e E_e (Q_{\beta^-_c} + m_e - E_e )^2 }{\displaystyle 1 - e^{\textstyle
-\,2\pi \alpha E_e /\sqrt{E^2_e - m^2_e}}}\nonumber\\
\hspace{-0.3in}&&\times\,\Big(1 + \frac{\alpha}{2\pi}\,g(E_e)\Big)\,dE_e =
0.0404\,{\rm MeV}^5,
\end{eqnarray}
where we have taken into account the contribution of the radiative
corrections \cite{RC1}. The function $g(E_e)$, calculated in \cite{RC1},
is given in the Appendix.

Neglecting the contribution of quadratic values of the scalar and
tensor couplings, the continuum-state $\beta^-$--decay rate of the
 neutron is
\begin{eqnarray}\label{label20}
\hspace{-0.3in}&&\tilde{\lambda}_{\beta^-_c} =\lambda_{\beta^-_c}\,(1
+ b\,\Delta_F),
\end{eqnarray}
where $\Delta_F = \tilde{f}^{(\gamma)}(Q_{\beta^-_c}, Z = 1)/
f^{(\gamma)}(Q_{\beta^-_c}, Z = 1) = 0.6612$ and $b$ is the Fierz term
\cite{NDR3} (see Eqs.(\ref{label30}) - (\ref{label32})) equal to
\begin{eqnarray}\label{label21}
b = 2\,\frac{g_S + 3g_A g_T}{1 + 3 g^2_A} = 0.0032(23).
\end{eqnarray}
The numerical value of the Fierz term is obtained from the fit of the
 experimental value $\tau^{\exp}_{\beta^-_c} = 878.5(8)\,{\rm s}$
 \cite{NDR2}. For the linear combination $g_S + 3 g_A g_T$ of the
 scalar and tensor coupling constant we get
\begin{eqnarray}\label{label22}
g_S + 3g_A g_T = 0.0094(70).
\end{eqnarray}
The contribution of the scalar and tensor weak interactions changes
the amplitude of the bound-state $\beta^-$--decay as follows
\begin{eqnarray}\label{label23}
\hspace{-0.3in}&&M(n\to {\rm H}^{(ns)} + \tilde{\nu}_e) = G_F
V_{ud}\sqrt{2 m_n 2 E_{\rm H}2 E_{\tilde{\nu}_e}}\nonumber\\
\hspace{-0.3in}&&\times\,\Big\{(1 + g_S)\,[\varphi^{\dagger}_e
\chi_{_{\tilde{\nu}_e}}][\varphi^{\dagger}_{p} \varphi_n] - (g_A +
g_T)\nonumber\\
\hspace{-0.3in}&&
\times\,[\varphi^{\dagger}_e\vec{\sigma}\,\chi_{_{\tilde{\nu}_e}}]\cdot
[ \varphi^{\dagger}_p\vec{\sigma}\,\varphi_n]\Big\}\,
\psi^*_{(ns)_F}(0).
\end{eqnarray}
The bound-state $\beta^-$--decay rate of the  neutron is equal to 
\begin{eqnarray}\label{label24}
\tilde{\lambda}_{\beta^-_b} &=& ((1 + g_S)^2 + 3(g_A +
g_T)^2)\,\zeta(3)\, G^2_F|V_{ud}|^2\nonumber\\ &&\times\,\frac{\alpha^3
m^3_e}{\pi^2}\,\sqrt{(m_p + m_e)^2 + Q^2_{\beta^-_c}}
\,\frac{Q^2_{\beta^-_c}}{m_n}.
\end{eqnarray}
Neglecting the contribution of the quadratic coupling constants of the
scalar and tensor weak interactions we get
\begin{eqnarray}\label{label25}
\tilde{\lambda}_{\beta^-_b} &=& (1 + b)\,\lambda_{\beta^-_b} =
\lambda_{\beta^-_b}.
\end{eqnarray}
Thus, the ratio $\tilde{R}_{b/c} = \tilde{\lambda}_{\beta^-_b}/
\tilde{\lambda}_{\beta^-_c}$ of the bound-state and continuum-state
$\beta^-$--decay rates of the  neutron is not changed
$\tilde{R}_{b/c} = 3.92\times 10^{-6}$.

\section{Helicity amplitudes and angular distributions of 
bound-state $\beta^-$--decay rates of neutron}

If the axis of the antineutrino--spin quantisation is inclined
relative to the axis of the neutron--spin quantisation with a polar
angle $\vartheta$, the wave function $\chi_{_{\tilde{\nu}_e}}$ can be
taken in the following form
\begin{eqnarray}\label{label26}
\chi_{_{\tilde{\nu}_e}} = {- e^{- i\varphi}\sin\frac{\vartheta}{2}\choose
\cos\frac{\vartheta}{2}},
\end{eqnarray}
 where $\varphi$ is an azimuthal angle.  The contributions of
different spinorial states to the helicity amplitudes of the
bound-state $\beta^-$--decay as functions of the angles $\vartheta$
and $\varphi$ are adduced in Table I.
{\renewcommand{\arraystretch}{1.5} \renewcommand{\tabcolsep}{0.13cm}
\begin{table}[h]
\begin{tabular}{|l|c|c|c|c|}
\hline $\sigma_n$ & $\sigma_p$ & $\sigma_e$ &
$\sigma_{\tilde{\nu}_e}$& $f$\\ \hline $+\frac{1}{2} $ & $+\frac{1}{2}
$ & $-\frac{1}{2} $ &$+\frac{1}{2} $ & $(1 + g_S + g_A +
g_T)\,\cos\frac{\vartheta}{2}$\\ \hline $+\frac{1}{2} $ &
$+\frac{1}{2} $ & $+\frac{1}{2} $ &$+\frac{1}{2} $ & $-(1 + g_S - g_A
- g_T)\,e^{\,- i\varphi}\sin\frac{\vartheta}{2}$ \\ \hline
$+\frac{1}{2} $ & $-\frac{1}{2} $ & $-\frac{1}{2} $ &$+\frac{1}{2} $ &
$0$ \\ \hline $+\frac{1}{2} $ & $-\frac{1}{2} $ & $+\frac{1}{2} $
&$+\frac{1}{2} $ & $- 2(g_A + g_T)\, \cos\frac{\vartheta}{2}$ \\
\hline $-\frac{1}{2} $ & $+\frac{1}{2} $ & $-\frac{1}{2} $
&$+\frac{1}{2} $ & $ 2 (g_A + g_T)\,e^{\,-
i\varphi}\sin\frac{\vartheta}{2}$ \\ \hline $-\frac{1}{2} $ &
$+\frac{1}{2} $ & $+\frac{1}{2} $ &$+\frac{1}{2} $ & $0$ \\ \hline
$-\frac{1}{2} $ & $-\frac{1}{2} $ & $-\frac{1}{2} $ &$+\frac{1}{2} $ &
$(1 + g_S - g_A - g_T)\,\cos\frac{\vartheta}{2}$ \\ \hline
$-\frac{1}{2} $ & $-\frac{1}{2} $ & $+\frac{1}{2} $ &$+\frac{1}{2} $ &
$-(1 + g_S + g_A + g_T)\,e^{\,- i\varphi}\sin\frac{\vartheta}{2}$ \\
\hline
\end{tabular}
\caption{The contributions of different spinorial states of the
interacting particles to the amplitudes of the bound-state
$\beta^-$--decay of the neutron and the antineutrino in the state with
the wave function Eq.(\ref{label26}); $f$ is defined by $f = (1 +
g_S)[\varphi^{\dagger}_e\chi_{_{\tilde{\nu}_e}}][\varphi^{\dagger}_p
\varphi_n] - (g_A + g_T)
[\varphi^{\dagger}_e\vec{\sigma}\,\chi_{_{\tilde{\nu}_e}}]\cdot [
\varphi^{\dagger}_p\vec{\sigma}\,\varphi_n]$. }
\end{table}

Using the results in Table 1 we get the helicity amplitudes $M(n \to
H_{FM_F}+ \tilde{\nu}_e)_{\sigma_n,+\frac{1}{2}}$
\begin{eqnarray}\label{label27}
&&M(n \to H_{00} + \tilde{\nu}_e)_{+\frac{1}{2},+\frac{1}{2}}
=\nonumber\\ &&= M_0 \frac{1 + 3 g_A + g_S + 3 g_T}{\sqrt{2}}\,
\cos\frac{\vartheta}{2},\nonumber\\ &&M(n \to H_{1,+1} +
\tilde{\nu}_e)_{+\frac{1}{2},+\frac{1}{2}} =\nonumber\\ &&= - M_0 (1 -
g_A + g_S - g_T) e^{\,- i\varphi}\,\sin\frac{\vartheta}{2},\nonumber\\
&&M(n \to H_{10} + \tilde{\nu}_e)_{+\frac{1}{2},+\frac{1}{2}} =
\nonumber\\ &&= M_0 \frac{1 - g_A + g_S - g_T}{\sqrt{2}}\,
\cos\frac{\vartheta}{2},\nonumber\\ &&M(n \to H_{1,-1} +
\tilde{\nu}_e)_{+\frac{1}{2},+\frac{1}{2}} = 0,\nonumber\\ &&M(n \to
H_{00} + \tilde{\nu}_e)_{-\frac{1}{2},+\frac{1}{2}} = \nonumber\\ && =
M_0 \frac{1 + 3 g_A + g_S + 3 g_T }{\sqrt{2}}\, e^{\,-
i\varphi}\sin\frac{\vartheta}{2},\nonumber\\ &&M(n \to H_{1,+1} +
\tilde{\nu}_e)_{-\frac{1}{2},+\frac{1}{2}} = 0,\nonumber\\ &&M(n \to
H_{10} + \tilde{\nu}_e)_{-\frac{1}{2},+\frac{1}{2}} = \nonumber\\ &&=
- M_0 \frac{1 - g_A + g_S - g_T}{\sqrt{2}}\,e^{\,-
i\varphi}\sin\frac{\vartheta}{2},\nonumber\\ &&M(n \to H_{1,-1} +
\tilde{\nu}_e)_{-\frac{1}{2},+\frac{1}{2}} = \nonumber\\ && = M_0(1 -
g_A + g_S - g_T)\,\cos\frac{\vartheta}{2}.
\end{eqnarray}
The angular distributions of the probabilities of the bound-state
$\beta^-$--decays of the polarised neutron are equal to
\begin{eqnarray}\label{label28}
\hspace{-0.3in}&&4\pi \frac{dR^{(+)}_{F =
0}}{d\Omega} = \frac{1}{8} \frac{(1 +
3g_A)^2}{1 + 3 g^2_A}\,\frac{1}{1 + b}\nonumber\\
\hspace{-0.3in}&&\times\,\Big(1 + 2\,\frac{g_S + 3g_T}{1 + 3
g_A}\Big)\,(1 + \cos\vartheta),\nonumber\\
\hspace{-0.3in}&&4\pi \frac{dR^{(-)}_{F =
0}}{d\Omega} = \frac{1}{8} \frac{(1
+ 3g_A)^2}{1 + 3 g^2_A}\,\frac{1}{1 + b}\nonumber\\
\hspace{-0.3in}&&\times\,\Big(1 + 2\,\frac{g_S + 3g_T}{1 + 3
g_A}\Big)\,(1 - \cos\vartheta),\nonumber\\
\hspace{-0.3in}&&4\pi \frac{dR^{(+)}_{F =
1}}{d\Omega} = \frac{1}{8} \frac{(1
- g_A)^2}{1 + 3 g^2_A}\frac{1}{1 + b}\nonumber\\
\hspace{-0.3in}&&\times\,\Big(1 + 2\,\frac{g_S - g_T}{1 - g_A}\Big)\,
(3 - \cos\vartheta),\nonumber\\\hspace{-0.3in}&& 4\pi
\frac{dR^{(-)}_{F = 1}}{d\Omega} =
\frac{1}{8} \frac{(1 - g_A)^2}{1 + 3 g^2_A}\,\frac{1}{1 +
b}\nonumber\\
\hspace{-0.3in}&&\times\,\Big(1 + 2\,\frac{g_S - g_T}{1 - g_A}\Big)\,
(3 + \cos\vartheta),
\end{eqnarray}
where $R^{(\pm)}_F =
(\lambda_{\beta^-_b})^{(\pm)}_F/\tilde{\lambda}_{\beta^-_b}$ and indices
$(\pm)$ stand for the polarisations of the neutron.

For $g_S = g_T = 0$ these angular distributions of the decay
probabilities agree well with those obtained by Song in
\cite{BS1}. Our polar angle $\vartheta$ is related to the polar angle
$\theta$ in Song's paper as $\vartheta = \pi - \theta$.

The angular distributions, given in Eq.(\ref{label28}), can be used
for the experimental search for the bound-state $\beta^-$--decay of
the polarised neutron into hydrogen in the hyperfine state with $F =
1$. Since in the directions $\cos\vartheta = \mp 1$ the angular
distributions of the probabilities of the production of hydrogen in
the hyperfine state with $F=0$ vanish, so for $\cos\vartheta = \mp 1$
one can detect only the bound-state $\beta^-$--decays of the neutron
into hydrogen in the hyperfine state with $F = 1$.

The probabilities of decays into hydrogen in the certain hyperfine
states are equal to
\begin{eqnarray}\label{label29}
\hspace{-0.3in}&&R_{F = 0} =\frac{(\lambda_{\beta^-_b})_{F =
 0}}{\lambda_{\beta^-_b}} = \frac{1}{4}\,\frac{(1 + 3g_A)^2}{1 + 3
 g^2_A}\frac{1}{1 + b}\nonumber\\
 \hspace{-0.3in}&&\times\,\Big(1 + 2\,\frac{g_S + 3g_T}{1 + 3
 g_A}\Big) = 0.987(2)\,\Big(1 + 2\,\frac{g_S + 3g_T}{1 + 3
 g_A}\Big),\nonumber\\
\hspace{-0.3in}&&R_{F = 1} = \frac{(\lambda_{\beta^-_b})_{F =
 1}}{\lambda_{\beta^-_b}} = \frac{3}{4}\,\frac{(1 - g_A)^2}{1 + 3
 g^2_A}\,\frac{1}{1 + b}\nonumber\\
 \hspace{-0.3in}&&\times\,\Big(1 + 2\,\frac{g_S - g_T}{1 - g_A}\Big) =
 0.010(0)\,\Big(1 + 2\,\frac{g_S - g_T}{1 - g_A}\Big),\nonumber\\
 \hspace{-0.3in}&&
\end{eqnarray}
where we have used the numerical values $g_A = 1.2750(9)$ and $b =
0.0032(23)$.

\section{Electron spectrum of continuum-state $\beta^-$--decay of neutron
 with correlation coefficients}

The experimental measurement of the value of the axial coupling
constant $g_A$ can be carried out by measuring the electron energy
spectrum and correlation coefficients \cite{NDR3}. The electron energy
spectrum of the continuum-state $\beta^-$--decay of the neutron is
equal to
\begin{eqnarray}\label{label30}
\hspace{-0.3in}&&\frac{d^5 \lambda^{(\gamma)}_{\beta^-_c}}{dE_e d\Omega_e
d\Omega_{\tilde{\nu}_e}} = (1 + 3g^2_A + g^2_S + 3
g^2_T)\nonumber\\
\hspace{-0.3in}&&\times\,\frac{G^2_F|V_{ud}|^2}{16\pi^5}\,(Q_{\beta^-_c}
+ m_e - E_e)^2 E_e\sqrt{E^2_e - m^2_e}\nonumber\\
\hspace{-0.3in}&&\times\,F(E_e, Z = 1)\,\Big(1 +
\frac{\alpha}{2\pi}\,g(E_e)\Big)\,\Big(1 + a\,\frac{\vec{k}_e\cdot
\vec{k}_{\tilde{\nu}_e}}{E_e E_{\tilde{\nu}_e}}\nonumber\\
\hspace{-0.3in}&&+ b\,\frac{m_e}{E_e} + A\,\frac{\vec{\xi}\cdot
\vec{k}_e}{E_e}+ B\,\frac{\vec{\xi}\cdot
\vec{k}_{\tilde{\nu}_e}}{E_{\tilde{\nu}_e}}\Big),
\end{eqnarray}
where the coefficients $a$, $A$ and $B$ define the correlations
between momenta of electron and antineutrino, neutron spin and
electron momentum, and neutron spin and antineutrino momentum,
respectively, $\vec{\xi}$ is the unit polarisation vector of the
neutron.  The Fierz term $b$ \cite{NDR3} describes a deviation from
the $V - A$ theory of weak interactions. The correlation coefficients
are equal to
\begin{eqnarray}\label{label31}
a &=&\frac{1 - g^2_A - g^2_S + g^2_T}{1 + 3 g^2_A + g^2_S + 3
g^2_T},\nonumber\\ b &=&\frac{2(g_S + 3 g_A g_T)}{1 + 3 g^2_A + g^2_S
+ 3 g^2_T},\nonumber\\A &=&-\,2\frac{g_A(g_A - 1) + g_T( g_S - g_T)}{1
+ 3 g^2_A + g^2_S + 3 g^2_T},\nonumber\\ B &=&+\,2\,\frac{ g_A(g_A +
1) + g_T(g_S + g_T)}{1 + 3 g^2_A + g^2_S + 3 g^2_T}\nonumber\\ &&+
2\,\frac{g_T + g_A (g_S + 2 g_T)}{1 + 3 g^2_A + g^2_S + 3
g^2_T}\,\frac{m_e}{E_e}.
\end{eqnarray}
Neglecting the contribution of $g^2_S$, $g^2_T$ and $g_Sg_T$ we get
\begin{eqnarray}\label{label32}
\hspace{-0.3in}a &=&\frac{1 - g^2_A}{1 + 3 g^2_A}\;,\; b =
2\,\frac{g_S + 3 g_A g_T}{1 + 3 g^2_A},\nonumber\\ \hspace{-0.3in}A
&=&-\,2\,\frac{g_A(g_A - 1))}{1 + 3 g^2_A}\;,\; B =
+\,2\,\frac{g_A(g_A + 1)}{1 + 3 g^2_A}\nonumber\\ \hspace{-0.3in}&& +
2\,\frac{g_T + g_A (g_S + 2 g_T)}{1 + 3 g^2_A }\,\frac{m_e}{E_e}.
\end{eqnarray}
The coefficients $a$ and $A$ agree well with the results adduced in
\cite{NDR3}, whereas the coefficient $B$ differs from that, given in
\cite{NDR3}, by the term inversely proportional to the energy of the
electron and linear in scalar and tensor coupling constants. The value
of the Fierz term $b = 0.0032(23)$ is given in Eq.(\ref{label21}).

\section{Numerical value of CKM matrix element $|V_{ud}|$ in $V-A$ theory 
of weak interactions}

For the calculation of the lifetime of the neutron we have used the
numerical value $|V_{ud}| = 0.97419(22)$ of the CKM matrix element,
proposed in \cite{PDG08}.

In this section we calculate the value of the CKM matrix element
$|V_{ud}|$ in the $V-A$ theory of weak interactions, using our
expression for the continuum-state $\beta^-$--decay rate of the
neutron Eq.(\ref{label11}), calculated for the axial coupling constant
$g_A = 1.2750(9)$ \cite{NDR3} and accounting for the radiative
corrections, and the experimental values of the lifetimes of the
neutron \cite{NDR1,PDG08}.  From Eq.(\ref{label11}) with
$f(Q_{\beta^-_c}, Z = 1) \to f^{(\gamma)} (Q_{\beta^-_c}, Z = 1)$ we
get
\begin{eqnarray}\label{label33}
\hspace{-0.3in}&&|V_{ud}|^2 =
\frac{4910.22}{\tau^{(\exp)}_{\beta^-_c}(1 + 3 g^2_A)}.
\end{eqnarray}
Using the experimental values of the lifetimes
$\tau^{(\exp)}_{\beta^-_c} = 878.5(8)\,{\rm s}$ and
$\tau^{(\exp)}_{\beta^-_c} = 885.7(8)\,{\rm s}$, measured in
\cite{NDR1} and \cite{NDR2}, respectively, we obtain
\begin{eqnarray}\label{label34}
\hspace{-0.3in}|V_{ud}| = \left\{\begin{array}{r@{\;,\;}l} 0.9752(7) &
\tau^{(\exp)}_{\beta^-_c} = 878.5(8)\,{\rm s}\\ 0.9713 (7)&
\tau^{(\exp)}_{\beta^-_c} = 885.7(8)\,{\rm s}.
\end{array}\right.
\end{eqnarray}
\begin{figure}
\centering
\includegraphics[height=0.29\textheight]{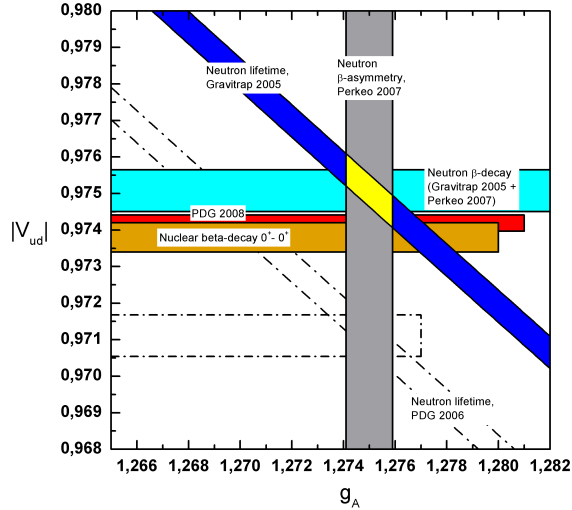}
\caption{The dependence of the CKM matrix element $|V_{ud}|$ on the
values of the lifetime of the neutron and the axial coupling constant
$g_A$.}
\end{figure}

In Fig.\,1 we show a dependence of the CKM matrix element on the
values of the lifetime of the neutron and the axial coupling constant
$g_A$. The yellow area shows that the value $|V_{ud}| = 0.9752(7)$,
calculated for the lifetime $\tau^{(\exp)}_{\beta^-_c} =
878.5(8)\,{\rm s}$, agrees with both $|V_{ud}| = 0.97419(22)$ and
$|V_{ud}| = 0.9738(4)$.

One can see that the value $|V_{ud}| = 0.9713(7)$, calculated for the
lifetime $\tau^{(\exp)}_{\beta^-_c} = 885.7(8)\,{\rm s}$, is ruled
out by the experimental value $|V_{ud}| = 0.9738(4)$, measured from
the superallowed $0^+ \to 0^+$ nuclear $\beta^-$--decays, caused by
pure Fermi transitions only \cite{NDR3,NDR5}, and the unitarity of the
CKM matrix elements giving $|V_{ud}| = 0.97419(22)$ \cite{PDG08}.

\section{Conclusive discussion}

 We have recalculated the continuum-state and bound-state
$\beta^-$--decay rates of the neutron. Taking into account the
contributions of weak and strong interactions for the lifetime of the
neutron we get the value $\tau_{\beta^-_c} = 914.8(1.2)\,{\rm s}$,
where the error $\pm 1.2\,{\rm s}$ is caused by the experimental error
of the axial coupling constant $g_A = 1.2750(9)$ and the CKM matrix
element $|V_{ud}| = 0.97419(22)$ \cite{PDG08}.  Including the
radiative corrections \cite{RC1,RC2}, the theoretical value of the
lifetime of the neutron changes to $\tau^{(\gamma)}_{\beta^-_c} =
880.4(1.1)\,{\rm s}$. It agrees well the experimental value
$\tau^{(\exp)}_{\beta^-_c} = 878.5(8)\,{\rm s}$ \cite{NDR1}.

We would like to accentuate that the radiative corrections are
universal and make up  about $3.9\,\%$. The theoretical value of the
radiative corrections, calculated in this paper
\begin{eqnarray}\label{label35}
R_{RC} = \frac{f^{(\gamma)}(Q_{\beta^-_c}, Z =
1)}{f(Q_{\beta^-_c}, Z = 1)} = 1.03912,
\end{eqnarray}
agrees well with the value $R_{RC} = 1.03886(39)$, given in
\cite{NDR3}. 

The agreement of the theoretical value of the lifetime of the neutron
$\tau_{\beta^-_c} = 880.4(1.1)\,{\rm s}$ with the experimental value
$\tau^{(\exp)}_{\beta^-_c} = 878.5(8)\,{\rm s}$, measured in
\cite{NDR1}, is fully due to the axial coupling constant $g_A =
1.2750(9)$ and the CKM matrix element $|V_{ud}| = 0.97419(22)$
\cite{PDG08}.

Using our expression (\ref{label11}) for the continuum-state
$\beta^-$--decay rate with the Fermi integral, accounting for the
contribution of radiative corrections, the axial coupling constant
$g_A = 1.2750(9)$ and the experimental lifetimes of the neutron
$\tau^{(\exp)}_{\beta^-_c} = 878.5(8)\,{\rm s}$ \cite{NDR1} and
$\tau^{(\exp)}_{\beta^-_c} = 885.7(8)\,{\rm s}$ \cite{PDG08} we got
the values of the CKM matrix element $|V_{ud}| = 0.9752(7)$ and
$|V_{ud}| = 0.9713(7)$, respectively.

It is seen that $|V_{ud}| = 0.9713(7)$ is ruled out by the values
$|V_{ud}| = 0.9738(4)$ and $|V_{ud}| = 0.97419(22)$, defined from the
superallowed $0^+ \to 0^+$ nuclear $\beta^-$--decays \cite{NDR3,NDR5}
and the unitarity condition for the CKM matrix elements \cite{PDG08},
respectively.  This implies that the nature singles out the lifetime
of the neutron $\tau^{(\exp)}_{\beta^-_c} = 878.5(8)\,{\rm s}$
\cite{NDR1}. Of course, this assertion should be confirmed by
experimental data in other terrestrial laboratories. Some hints of the
validity of this assertion can be found also in cosmology
\cite{BAU1,BAU2}.

For the axial coupling constant $g_A = 1.2750(9)$ the correlation
coefficients are equal to
\begin{eqnarray}\label{label36}
\hspace{-0.5in}&&a^{\rm (th)} = -0.1065(3)\;,\; a^{(\exp)} =
-0.103(4),\nonumber\\
\hspace{-0.5in}&&B^{\rm (th)} = + 0.9871(4) + 2\,\frac{g_T + g_A(g_S +
2 g_T)}{1+ 3g^2_A}\,\frac{m_e}{E_e},\nonumber\\
\hspace{-0.5in}&&B^{(\exp)} = +
0.9821(40),\nonumber\\
\hspace{-0.5in}&&C^{(\rm (th)} = - 0.2385(1)\;,\; C^{(\exp)} = -
0.2377(26),
\end{eqnarray}
where the coefficient $B^{(\exp)}= + 0.9821(40)$ has been measured in
\cite{SAA1,SAA2}, $C = - 0.27484\,(A + B)$ is the proton asymmetry,
measured in \cite{NDR4}.

We remind that the value $g_A = 1.2750(9)$ of the axial coupling
constant has been calculated from the fit of the experimental value of
the neutron spin--electron correlation coefficient $A^{(\exp)} =
-\,0.11933(34)$, which has been obtained in \cite{NDR3} as an averaged
value over PERKEO II measurements \cite{PERII1,PERII2}.

The deviation of the theoretical value of the lifetime of the free
neutron $\tau^{\rm (th)}_{\beta^-_c} = 880.1(1.1)\,{\rm s}$ from the
experimental one $\tau^{\exp}_{\beta^-_c} = 878.5(8)\,{\rm s}$
\cite{NDR1} allows to take into the contributions of scalar and tensor
weak interactions, which can be added to the standard $V - A$
baryon--lepton weak interactions with coupling constants $g_S$ and
$g_T$, respectively. From the fit of the experimental value of the
lifetime of the neutron $\tau^{\exp}_{\beta^-_c} = 878.5(8)\,{\rm s}$
\cite{NDR1} we have found $g_S + 3 g_A g_T = 0.0094(70)$, caused by
the value of the Fierz term $b = 0.0032(23)$ Eq.(\ref{label21}).

Since standard $V - A$ weak interactions describe well the
experimental data on the coefficient of the neutron spin--antineutrino
momentum correlation, we set zero the contribution of the
energy--dependent term in the coefficient $B$. This gives $g_T +
g_A(g_S + 2 g_T) = 0$. Solving this equation together with the Fierz
term Eq.(\ref{label21}) we estimate the scalar and tensor coupling
constants
\begin{eqnarray}\label{label37}
g_S&=& +\,\frac{b}{2}\,\frac{(1 + 2 g_A)(1 + 3 g^2_A)}{1+ 2 g_A - 3
g^2_A} = -\,0.0251(181),\nonumber\\ g_T&=& -\,\frac{b}{2}\,\frac{g_A (1
+ 3 g^2_A)}{1+ 2 g_A - 3 g^2_A} = +\,0.0090(65)
\end{eqnarray}
The deviations from these values can be obtained experimentally by
measuring the neutron spin--antineutrino momentum correlation and the
bound-state $\beta^-$--decay rates of the neutron into hydrogen in
certain hyperfine states. As we have shown the measurement of the
angular distributions of the probabilities of the bound-state
$\beta^-$--decay of the polarised neutron into hydrogen in the
hyperfine states with $F = 1$ can be carried out at $\cos\vartheta =
\pm 1$.

Our angular distributions for the probabilities of the bound-state
$\beta^-$--decay rates of the neutrino into hydrogen in the certain
hyperfine state agree at $g_S = g_T = 0$ with those obtained by Song
\cite{BS1}.

\section{Appendix A: Radiative corrections to  continuum-state 
$\beta^-$--decay rate of  neutron}
\renewcommand{\theequation}{A-\arabic{equation}}
\setcounter{equation}{0}

Below we calculate the radiative corrections to the continuum-state
$\beta^-$--decay rate following the results obtained in
\cite{RC1,RC2}. Following \cite{RC1} we determine the continuum-state
$\beta^-$--decay rate with radiative corrections as follows
\begin{eqnarray}\label{labelA.1}
\hspace{-0.3in}\lambda^{(\gamma)}_{\beta^-_c} = (1 + 3 g^2_A)\,
    \frac{G^2_F |V_{ud}|^2}{2 \pi^3}\,f^{(\gamma)}(Q_{\beta^-_c}, Z =
    1),
\end{eqnarray}
where the Fermi integral $f^{(\gamma)}(Q_{\beta^-_c}, Z = 1)$ is
given  by
\begin{eqnarray}\label{labelA.2}
\hspace{-0.3in}&&f^{(\gamma)}(Q_{\beta^-_c}, Z = 1) = \nonumber\\
\hspace{-0.3in}&& = \int^{Q_{\beta^-_c} + m_e}_{m_e} \frac{2\pi\alpha
E^2_e(Q_{\beta^-_c} + m_e - E_e )^2 }{\displaystyle 1 - e^{\textstyle
-\,2\pi \alpha E_e /\sqrt{E^2_e - m^2_e}}}\nonumber\\
\hspace{-0.3in}&&\times\,\Big(1 + \frac{\alpha}{2\pi}\,g(E_e)\Big)\,dE_e =
0.0611\,{\rm MeV}^5.
\end{eqnarray}
The function $g(E)$, calculated in \cite{RC1,RC2}, is
\begin{eqnarray}\label{labelA.3}
\hspace{-0.3in}&&g(E_e) = 3 \,\Big[{\ell n}\Big(\frac{m_p}{m_e}\Big) -
    \frac{1}{4}\Big] + 4\,\Big[\frac{E}{2 \sqrt{E^2_e -
    m^2_e}}\nonumber\\\hspace{-0.3in} &&\times \,{\ell
    n}\Big(\frac{E_e + \sqrt{E^2 - m^2_e}}{E - \sqrt{E^2 -
    m^2_e}}\Big) - 1\Big]\Big[\frac{ Q_{\beta^-_c} + m_e - E_e}{3
    E_e}\nonumber\\
    \hspace{-0.3in}&& - \frac{3}{2} + {\ell n}\Big(\frac{2(
    Q_{\beta^-_c} + m_e - E_e)}{m_e}\Big)\Big] + \frac{4 E_e}{\sqrt{E^2_e -
    m^2_e}}\nonumber\\\hspace{-0.3in} &&\times
    \,F\Big(\frac{2\sqrt{E^2_e - m^2_e}}{E_e + \sqrt{E^2_e - m^2_e}}\Big) +
    \frac{E_e}{2 \sqrt{E^2_e - m^2_e}}\nonumber\\
    \hspace{-0.3in}&&\times\,{\ell n}\Big(\frac{E_e + \sqrt{E^2_e -
    m^2_e}}{E_e - \sqrt{E^2_e - m^2_e}}\Big)\Big[2\Big(1 + \frac{E^2_e -
    m^2_e}{E^2_e}\Big)\nonumber\\
    \hspace{-0.3in}&&+ \frac{ (Q_{\beta^-_c} + m_e - E_e)^2}{6 E^2_e} -
    2\,{\ell n}\Big(\frac{E_e + \sqrt{E^2_e - m^2_e}}{E_e - \sqrt{E^2_e -
    m^2_e}}\Big)\Big]\nonumber\\
    \hspace{-0.3in}&& + \Big[3 {\ell n}\Big(\frac{m_W}{m_p}\Big) +
    {\ell n}\Big(\frac{m_W}{m_{a_1}}\Big) + 4 
    {\ell n}\Big(\frac{m_Z}{m_W}\Big) + \frac{9}{4}\Big].\nonumber\\
    \hspace{-0.3in}&&
\end{eqnarray}
The numerical value of the Fermi integral is calculated for $m_W =
80.4\,{\rm GeV}$ and $m_Z = 90.2\,{\rm GeV}$, the masses of the $W$
and $Z$ bosons of the standard electroweak theory by Weinberg--Salam
\cite{PDG08}, and $m_{a_1} = 1.2\,{\rm GeV}$, the mass of the axial
meson \cite{PDG08,RC1}. The Spence function $F(x)$ is defined by
\begin{eqnarray}\label{labelA.4}
F(x) = \int^x_0 \frac{{\ell n}(1 - t)}{t}\,dt.
\end{eqnarray}
Following \cite{RC1} we have neglected the contributions of
electromagnetic corrections of order of $O(\alpha^2)$ and higher as
well as the contributions of the finite radius of the proton, which is
of order of $r_p \simeq 0.875(7)\,{\rm fm}$ \cite{PDG08}. This
approximation can be justified by using the results obtained in
\cite{RC2}. For $\tilde{f}^{(\gamma)}(Q_{\beta^-_c}, Z = 1)$ we get
$\tilde{f}^{(\gamma)}(Q_{\beta^-_c}, Z = 1) = 0.0404\,{\rm MeV}^5$.

The continuum-state $\beta^-$--decay rate of the neutron, accounting
for the radiative corrections, is $\lambda^{(\gamma)}_{\beta^-_c} =
1.1359(14)\times 10^{-3}\,{\rm s^{-1}}$. The theoretical value of the
lifetime of the neutron equal to $\tau^{(\gamma)}_{\beta^-_c} =
880.4(1.1)\,{\rm s}$ agrees well the experimental one
$\tau^{(\exp)}_{\beta^-_c} = 878.5(8)\,{\rm s}$ by Serebrov {\it et
al.} \cite{NDR1}. The error margins $\pm 1.1\,{\rm s}$ are fully
determined by the experimental error margins of the axial coupling
constant $g_A = 1.2750(9)$ \cite{NDR3} and the CKM matrix element
$|V_{ud}| = 0.97419(22)$ \cite{PDG08}.

\end{document}